# Potassium fluoride doped LaOFeAs multi-band superconductor: Evidence of extremely high upper critical field

S. Singh, D. Srikala, A. K. Singh and S. Patnaik
School of Physical Sciences, Jawaharlal Nehru Univeristy, New Delhi 110067 India

J. Prakash and A. K. Ganguli
Department of Chemistry, Indian Institute of Technology, Delhi, New Delhi 110016 India

Corresponding author

ashok@chemistry.iitd.ernet.in
spatnaik@mail.jnu.ac.in

# Abstract

The recently discovered superconductors based on oxypnictides have rekindled the search for new non-copper based high $T_c$ superconductors. After the initial report by Kamihara et al.[1] of superconductivity in La-O/F-Fe-As based compounds, there have been several reports on these new superconductors but the synthesis of pure phases of these oxypnictides has remained a challenge. Further, the recent high – field transport measurements[2] have established multi - band superconductivity in these oxypnictides. Here we describe a new methodology of synthesizing these oxypnictide superconductors with the commonly available potassium fluoride (KF) as a source of fluorine instead of the expensive $LaF_3$. This route also allows the substitution of potassium at lanthanum sites which leads to an increase in the upper critical field parameters as would be expected for a multi-band superconductor. We also report the highest reported $T_c$ (onset) of 28.50 K (± 0.05 K) and highest upper critical field at ambient pressure in the family of La – based oxypnictides. This new strategy of simultaneous hole and electron doping in the La-O layer shows the possibility of further optimizing the $T_c$ and $H_{c2}$ of this newest superconductor via a proper choice of dopants.

**Introduction**

In early 2006, superconductivity was discovered in an unexpected structure by doping fluorine in an oxypnictide LaOFeP[3] with a $T_c$ around 5 K and subsequent studies improved the $T_c$ to 26 K in a related As – derivative (LaOFeAs)[1]. The parent compound, LaOFeAs crystallizes in a simple tetragonal structure[4] and consists of alternating LaO and FeP layers and is an antiferromagnetic semi – metal which is reminiscent of copper oxide based high Tc superconductors[5]. It shows a spin density wave correlation at ~ 150 K[6] while doping with fluorine suppresses the magnetic instability and leads to superconductivity at 26 K[1]. The FeAs layers provide the conducting pathway and the LaO layers act as charge reservoirs[7]. La has been replaced by other rare-earths such as Nd, Pr, Gd, Ce and Sm to yield new superconductors with higher transition temperatures[8-12]. On the other hand the transition temperature is lowered on substituting Fe by Ni [13]. The highly reactive nature of metals and metal fluorides coupled with the toxicity of pnictogens (P and As) has restricted the ability to carry out reactions with gay abandon. Hence most of the reports on the synthesis have revolved around similar lines with the use of metals and $LnF_3$ (Ln = rare-earth) as the source of fluorine. Moreover, it is now well established that, both the upper critical field, the maximum field upto which superconductivity can be sustained and the irreverisibility field that corresponds to disappearance of bulk supercurrent denisty, can be tuned by controlling the intra and inter band scattering mechanism of a multiband superconductor[14,15]. Taking both these facts into consideration, we have attempted to design a simpler process to obtain fluorine doped - oxypnictides and to achieve improved superconducting properties. We describe a new methodology of synthesizing these oxypnictides with the commonly available KF as a source of fluorine instead of the expensive $LaF_3$ as used in all earlier reports[1,8-13]. The quality of the doped and undoped samples is confirmed from the sharp transitions, low room temperature resistivity, and high onset temperatures. KF doping also allows the formation of K-substituted superconducting compounds. We show that a substantial enhancement of the upper critical field can be achieved by simultaneous doping of potassium and fluorine at lanthanum and oxygen sites respectively.

**Experimental**

For the synthesis of $La_{1-x}K_xO_{1-x}F_xFeAs$ (x = 0.15 and 0.20), stoichiometric amounts of $La_2O_3$, FeAs, La and KF were sealed in evacuated silica ampoules ($10^{-4}$ torr) and heated at 1000 C for 48h . The powder was compacted(5 tonnes) and the disks were wrapped in Ta foil, sealed in evacuated silica ampoules and heated at 1180 C for 48 h. $La_{1.03}O_{0.9}F_{0.2}FeAs$ (K – free superconductors) was obtained by initially loading La, $La_2O_3$, FeAs, $LaF_3$ in a stoichiometric ratio corresponding to the composition $LaO_{0.9}F_{0.1}FeAs$. The mixture was heated at 950 C and then at 1000 C for 8 h. Additional $LaF_3$ was added such that the final composition corresponds to $La_{1.03}O_{0.9}F_{0.2}FeAs$. The resulting mixture was again heated as above at 1180 C for 48 h. Powder X-ray diffraction patterns of the finely ground powders were recorded with Cu-Kα radiation in the 2θ range of 10° to 70°. The lattice parameters were obtained from a least squares fit to the observed *d* values.

The resistivity measurements were carried out (I = 10mA) using a Cryogenic 8T Cryogen-free magnet in conjunction with a variable temperature insert (VTI). The samples were cooled in helium vapor and the temperature of the sample was measured with an accuracy of 0.05 K using a calibrated Cernox sensor wired to a Lakeshore 340 temperature controller. The external magnetic field ranging from 0- 6 Tesla was applied perpendicular to the probe current direction and the data were recorded during the warming cycle with the heating rate of 1K/min. The inductive part of the magnetic susceptibility was measured via a tunnel diode based rf penetration depth technique[16]. The sample was kept inside an inductor that formed a part of LC circuit of an ultrastable oscillator (~2.3MHz). A change in the magnetic state of the sample results in a change in the inductance of the coil and is reflected as a shift in oscillator frequency which is measured by an Agilent 53131A counter.

## Results and Discussions

Powder X-ray diffraction patterns of the nominal compositions of $LaO_{0.9}F_{0.1}FeAs$ after heating at 950 and 1000C are shown in Figs.1a and 1b. After heating at 950 C (Fig. 1a) unreacted $La_2O_3$ and FeAs were present along with 60% of the desired oxypnictide phase which increased to nearly 75% on further heating at 1000 C. On subsequent heating at 1180 C (with addition of $LaF_3$ resulting in nominal composition of $La_{1.03}O_{0.9}F_{0.2}FeAs$) led to an increase (~85%) in the oxypnictide phase (Fig. 1c). It may be noted that we had around 10 % of LaOF and 5% of $Fe_2As$ as secondary phases. In all earlier reports[1,17], secondary phases like LaOF, FeAs, $Fe_2As$, $La_{4.67}(SiO4)_3O$ and $La_2O_3$ have been observed. There is no report so far of the synthesis of a pure phase of La(O/F)FeAs.

The KF doped compositions, $La_{1-x}K_xO_{1-x}F_xFeAs$ (x = 0.15 and 0.2) were also heated at 1180 C. Powder XRD studies show that majority of the observed reflections could be satisfactorily indexed based on the tetragonal LaOFeAs (space group P4/nmm) phase, barring some weak reflections that were assigned to secondary phases (LaOF and $Fe_2As$). The refined lattice parameters (**a** = 4.0270(7)Å and **c** = 8.718(3) Å) for F - doped LaOFeAs were smaller than those reported for pure LaOFeAs[2] (**a** = 4.038Å, **c** = 8.753 Å), as expected from ionic size considerations. The first study of a F - doped compound ($LaO_{0.95}F_{0.05}FeAs$) showed lattice parameters of **a** = 4.0355 and **c** = 8.7393 Å[1].

In the KF doped compositions we observed the major phase to be tetragonal with **a** = 4.0293 and **c** = 8.718 Å for x = 0.15 and **a** = 4.039 and **c** = 8.672Å for the x = 0.2 phase. The increase in **a** lattice parameter (though marginal) is indicative of K doping since all reports (including ours) on fluorine doped compositions show a decrease in both **a** and **c** lattice parameters compared to the undoped LaOFeAs. There is a significant variation of the **c**-parameter (Δ**c** = 0.08 Å) among the different samples. The **a** – parameter has a much lower dependence (Δ**a** = 0.01 Å) on either F or K substitution. This implies a considerable phase - width in these compounds which has potential implication for obtaining optimized transition temperatures in this family of superconductors.

Figure 2 shows the evolution of the resistivity plots of at different sintering temperatures (950 – 1180 C). The plots marked a and b, correspond to $LaO_{0.9}F_{0.1}FeAs$ whose x-ray patterns are shown in Fig 1a and 1b, while the plot 'c' corresponds to $La_{1.03}O_{0.9}F_{0.2}FeAs$ (XRD pattern :1c). We find that the phase obtained by sintering at 950

C yields the non-superconducting state with characteristic decrease in resistivity at ~150 K corresponding to the spin density wave (SDW) transition[6]. On sintering at 1000 C, the high temperature region of the resistivity plot becomes much more linear and the SDW transition disappears which is indicative of incorporation of fluorine. The superconducting transition is achieved in $La_{1.03}O_{0.9}F_{0.2}FeAs$, the sample sintered at 1180 °C. The absolute value of resistivity at 300 K of the above sample is equal to 2 mΩ cm (also confirmed in the Van der pau geometry) which is the smallest value reported so far[1,17].

The zero field resistivity as a function of temperature for $LaO_{0.9}F_{0.2}FeAs$, $La_{0.85}K_{0.15}FeAsO_{0.85}F_{0.15}$ and $La_{0.8}K_{0.2}FeAsO_{0.8}F_{0.2}$ is depicted in Fig. 3. The insets in each panel show the resistivity to high temperatures and also the inductive part of susceptibility attesting the onset of bulk diamagnetic behavior. We note that the onset of superconductivity in $LaO_{0.9}F_{0.2}FeAs$ occurs at 28.50 K and 50% of the transition is achieved in a temperature interval of 0.9 K. So far this is the highest $T_c$ that has been reported in this series of compounds (La-O/F-Fe-As)[1,17,18]. The criteria used for the $T_c$ determination is the same as that has been used elsewhere[17] and is schematically elucidated in Fig. 3a. The onset of the superconducting transition in $La_{0.85}K_{0.15}FeAsO_{0.85}F_{0.15}$ and $La_{0.8}K_{0.2}FeAsO_{0.8}F_{0.2}$ occurs at 26.20 K and 26.45 K respectively (Fig 3b and 3c). The suppression of $T_c$ by doping of K in La sites in these oxypnictides appears to have similar origins as seen with Al and C doping in multiband $MgB_2$[19]. The high residual resistivity (Fig. 3a) ratio (RRR = $\rho_{300\ K} / \rho_{30\ K}$) for $LaO_{0.9}F_{0.2}FeAs$, implies that the phase is much more homogeneous as compared to those reported earlier[1]. We find a RRR value of 4.45 and K doping lowers it further suggesting enhanced impurity scattering. It is also interesting to note that while F acts as an electron donor and substitution of K in La sites leads to hole injection onto the FeAs layers, the overall transition temperature goes down marginally with K doping ( ~2 K).

We next discuss the transport properties in the presence of an external magnetic field. The in-field resistive transitions for the samples (a) $LaO_{0.9}F_{0.2}FeAs$ and (b) $La_{0.8}K_{0.2}FeAsO_{0.8}F_{0.2}$ are shown in Fig. 4. At an applied field of 6 Tesla (Fig 4a) the offset of the transition shifts by only ~ 4 K which is indicative of strong flux pinning. This suggests the as - grown defects to be effective pinning centers. The broadening of in-field resistive transition is more muted as compared to that of the cuprate superconductors and augurs well from the point of view of applications. The insets of Fig. 4 show the upper critical field ($H_{c2}$), and the irreversibility field ($H^*$) as a function of temperature obtained magnetic field dependent resistivity studies. We have used the 90% of $\rho_n$ (normal state resistivity at T = $T_c$) criteria[20] to define $H_{c2}$ and 10% criteria for the corresponding H*. The slope $dH_{c2}/dT$ is estimated to be -5.3 T/K which increases to -6.7 T/K with potassium doping. There is a marked curvature at low fields in the $H_{c2}$ versus T plot (Fig. 4b) which is another signature of a multi - band scattering mechanism[15]. We have therefore chosen the data between 2 T and 6T for the calculation of $dH_{c2}/dT$. The extrapolated $H_{c2}$ (0) values using the Werthamer – Helfand - Hohenberg (WHH) formula[20], $H_{c2}(0) = -0.693T_c (dH_{c2}/dT)$ yields an estimate of zero temperature upper critical field, $H_{c2}(0)$ ~105 T for $LaO_{0.9}F_{0.2}FeAs$ and ~122 T for $La_{0.8}K_{0.2}FeAsO_{0.8}F_{0.2}$. This is an enormous increase in $H_{c2}(0)$ compared to the highest value of 60 T reported earlier[2]. Due to the polycrystalline nature of the samples, this value of the upper critical field should be taken as $H_{c2}^{\parallel}$. The increase in the upper critical field is accompanied by a marginal decrease in $T_c$ with doping of 'K', which again is the hallmark of multiband superconductivity. From the $H_{c2}(0)$ values we estimate the mean field Ginzburg-Landau coherence length ($\xi_{GL} = (\Phi_0 / 2\pi H_{c2})^{1/2}$, $\Phi_0$ being the flux quantum = $2.07 \times 10^{-7}$ G cm$^2$) for these oxypnictide superconductors to be ~17 Å. These values are smaller as compared to those reported earlier (33 Å[17]).

In conclusion, we provide a novel synthesis route to prepare the lanthanum-based oxypnictide superconductor ($T_c$ = 26.45 K) using an inexpensive and easily available fluorinating agent, KF. The use of KF also allows potassium doping in La-O layers that leads to optimization of upper critical field parameters in this multiband superconductor. A variation (anisotropic) in lattice parameters is observed suggesting a definite phase width in these compositions. More importantly, we observe the highest $T_c$ of 28.50($\pm$0.05 K) in $LaO_{0.9}F_{0.2}FeAs$, and the highest upper critical field of 122 T ( by WHH extrapolation) in $La_{0.8}K_{0.2}FeAsO_{0.8}F_{0.2}$ reported so far in La-based oxypnictides. The resistive broadening in the presence of external magnetic field indicates robust flux pinning in these high field superconductors.

**Acknowledgement**

AKG and SP thank DST, Govt. of India financial support . AKG thanks IIT Delhi for infrastructurural support. JP thanks CSIR, Govt. of India for a fellowship.

**Figure captions**

**Figure 1.** Powder X-ray diffraction patterns of the La – based oxypnictides. (a) $LaO_{0.9}F_{0.1}FeAs$ sintered at 950 C, (b) at 1000 C, (c) $La_{1.03}O_{0.9}F_{0.2}FeAs$ sintered at 1180 C, (d) $La_{0.15}K_{0.15}O_{0.15}F_{0.15}$ FeAs sintered at 1180 C, and (e) $La_{0.8}K_{0.2}O_{0.2}F_{0.2}FeAs$ sintered at 1180 C. The impurity phases are marked by the letters shown in parenthesis, LaOF (O), FeAs (F) and $La_2O_3$ (L).

**Figure 2.** Plot of resistivity as a function of temperature for F-doped LaOFeAs.

**Figure 3.** Plot of resistivity and the real part of the magnetic susceptibility. (a)$La_{1.03}O_{0.9}F_{0.2}FeAs$, (b) $La_{0.85}K_{0.15}O_{0.85}F_{0.15}FeAs$ and (c) $La_{0.8}K_{0.2}O_{0.8}F_{0.2}FeAs$ superconductors. The insets in each panel show the corresponding resistivity upto room temperature and rf susceptibility. The $T_c$'s are found to be (a) 28.50, (b) 26.20 and (c) 26.45 respectively.

**Figure 4.** Plot of resistivity versus temperature in the presence of a magnetic field. (a) $La_{1.03}O_{0.9}F_{0.2}FeAs$ and (b) $La_{0.8}K_{0.2}O_{0.8}F_{0.2}FeAs$ superconductors. Insets show upper critical field (■) and irreversibility field (●) as a function of temperature.

**Fig. 1**

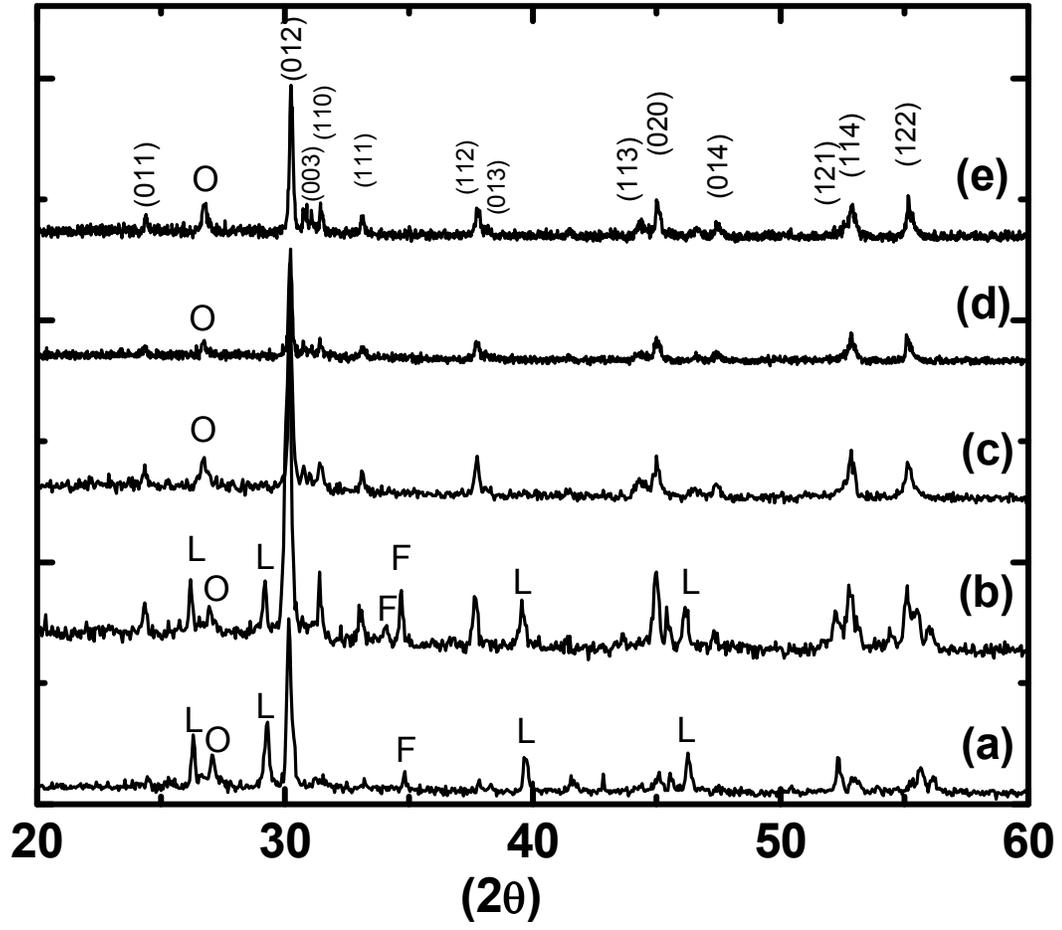

Fig. 2

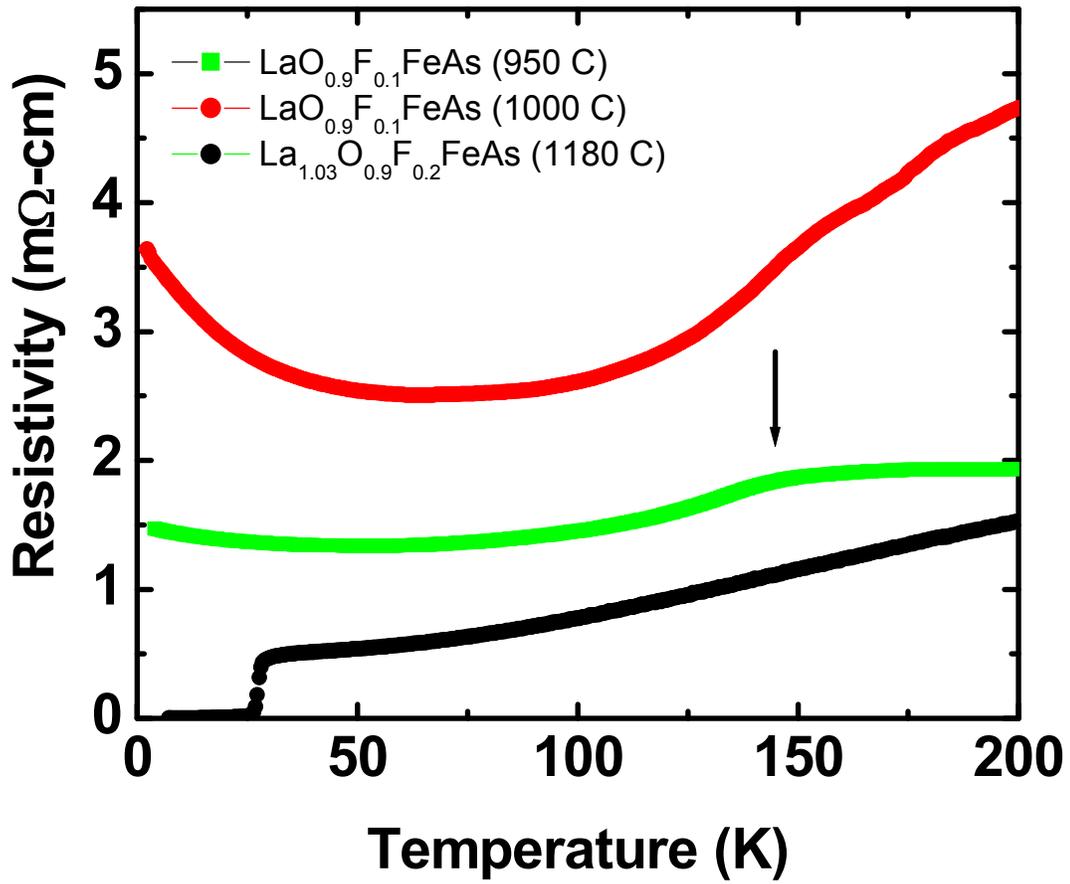

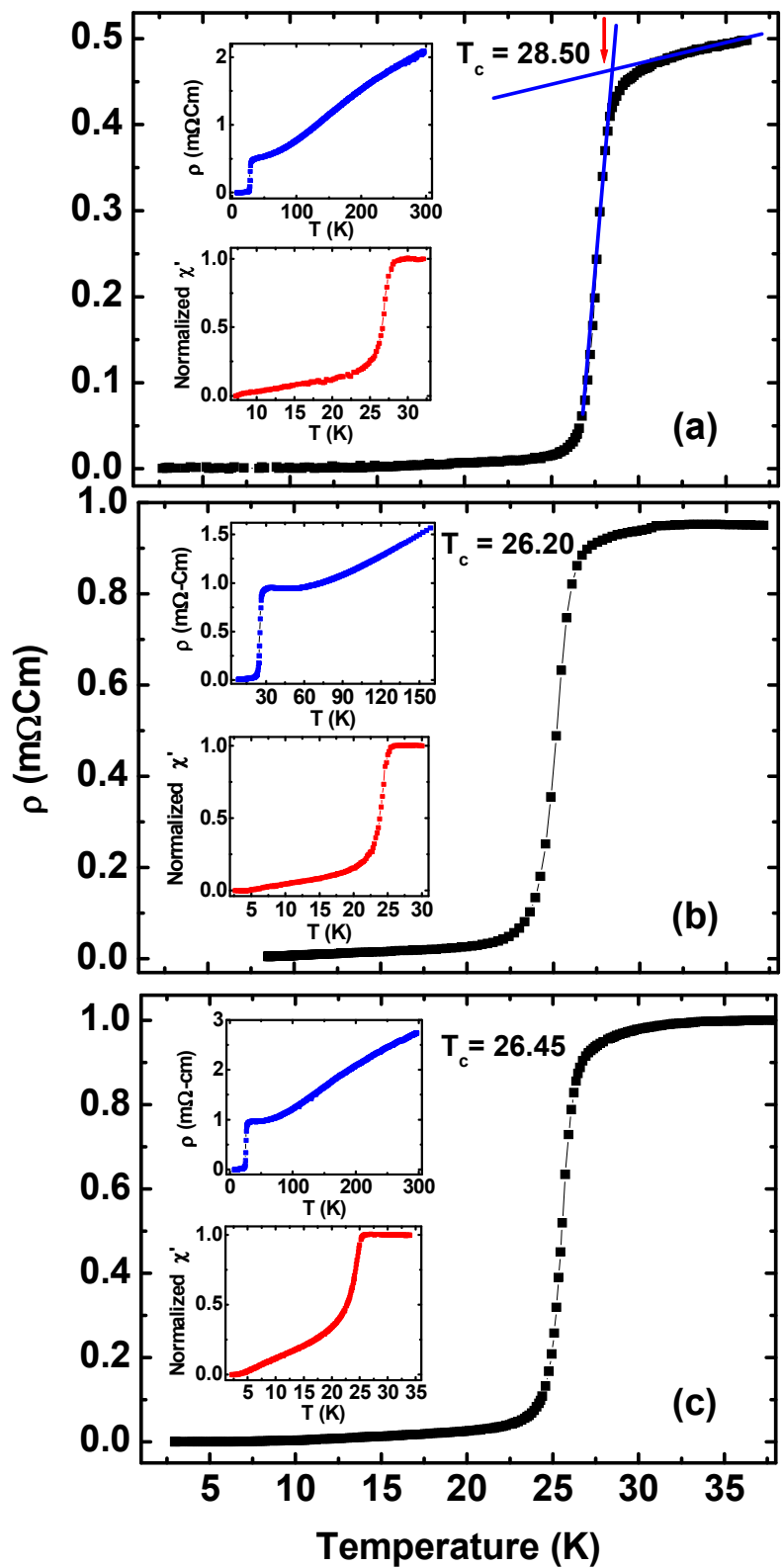

**Fig. 4a.**

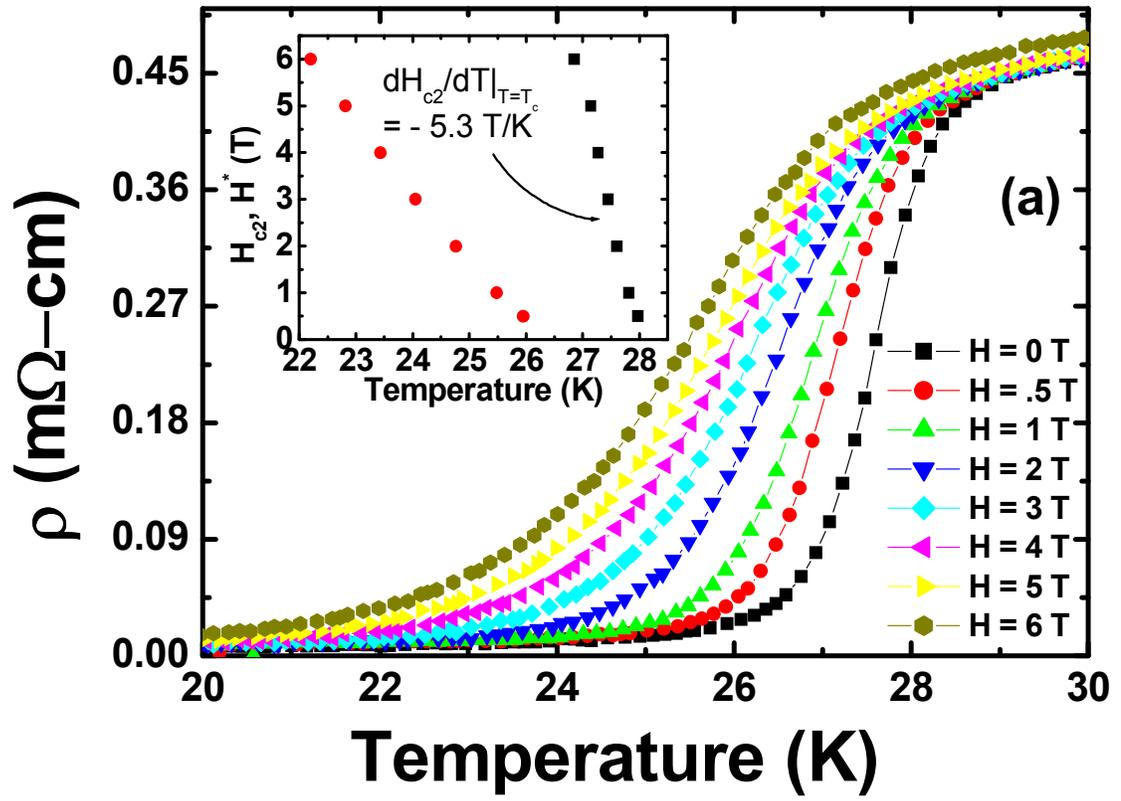

**Fig 4b**

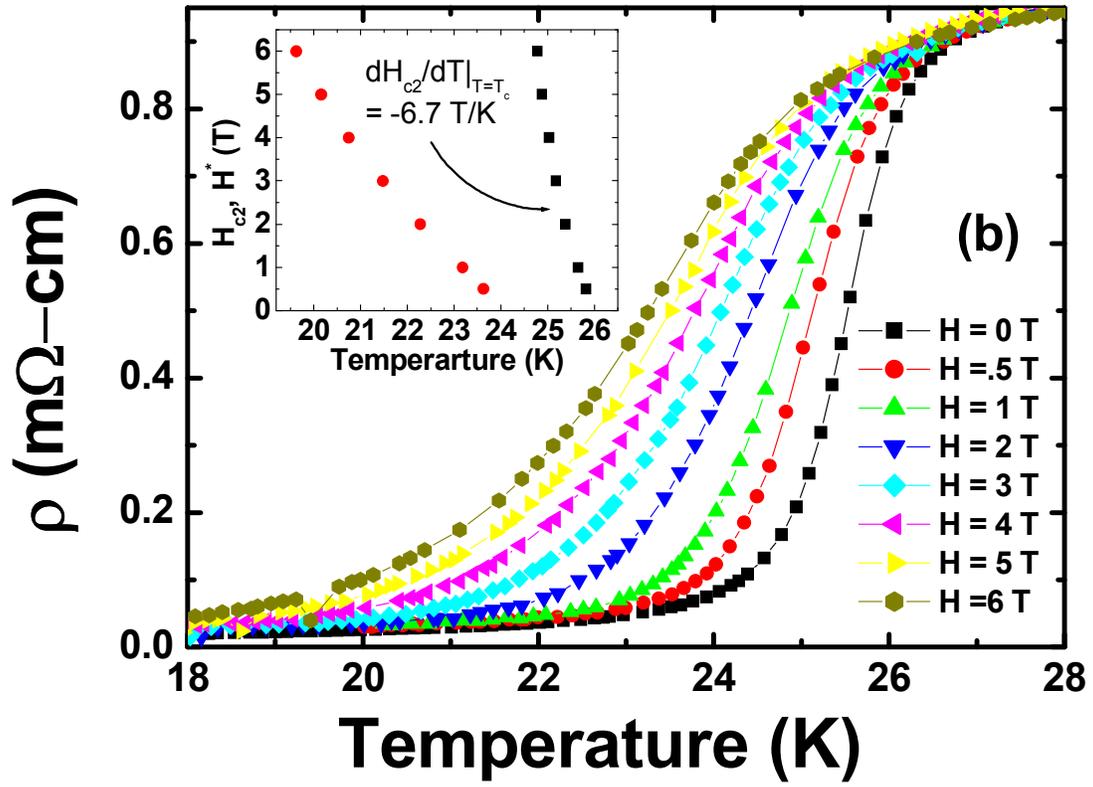